\documentclass{svmult}

\usepackage{makeidx}         
\usepackage{graphicx}        
\usepackage{multicol}        
\usepackage[bottom]{footmisc}

\makeindex             

\begin{document}

\title*{Change of ownership networks in Japan}
\author{
Wataru Souma\inst{1}\and
Yoshi Fujiwara\inst{2}\and
Hideaki Aoyama\inst{3}}
\institute{
ATR Network Informatics Laboratories, Kyoto 619-0288, Japan. \texttt{souma@atr.jp}\and
ATR Network Informatics Laboratories, Kyoto 619-0288, Japan. \texttt{yfujiwar@atr.jp}\and
Department of Physics, Graduate School of Science, Kyoto University, Yoshida,
Kyoto 606-8501, Japan. \texttt{aoyama@phys.h.kyoto-u.ac.jp}}
\maketitle

\vspace{5mm}
{\bfseries Summary.}
As complex networks in economics, we consider Japanese shareholding networks as they existed
in 1985, 1990, 1995, 2000, 2002, and 2003.
In this study, we use as data lists of shareholders
for companies listed on the stock market or on the
over-the-counter market. 
The lengths of the shareholder lists vary
with the companies, and we use lists for the top 20 shareholders.
We represent these shareholding networks as a directed graph by drawing arrows from shareholders
to stock corporations.
Consequently, the distribution of incoming edges has an upper bound, while that of outgoing edges
has no bound. This representation shows that for all years the distributions of outgoing degrees can be
well explained by the power
law function with an exponential tail. The exponent depends on the year and the country,
while the power law shape is maintained universally. We show that the exponent
strongly correlates with the long-term shareholding rate and the cross-shareholding rate.

\vspace{5mm}

\section{Introduction}
\label{sec:1}
Recently, many studies have revealed the true structure of real-world networks \cite{barabasi2002,dm2003}.
This development also holds true in the field of econophysics. Such studies have investigated
business networks \cite{sfa2003},
shareholding networks \cite{sfa2004, sfa2005a, sfa2005b, gbcsc2003},
world trade networks \cite{ljc2003,ljc2004},
and corporate board networks \cite{bbw2003,dyb2003}.

By common practice, if we intend to discuss networks, we must define
their nodes and edges. Edges represent the relationships between nodes.
The subject of this study is the ownership network.
Accordingly, we consider companies as nodes and the shareholding relationships
between them as edges.

In this article, we consider Japanese shareholding networks as they existed
in 1985, 1990, 1995, 2000, 2002, and 2003
(see Ref.~\cite{gbcsc2003} for shareholding
networks in MIB, NYSE, and NASDAQ).
We use data published by Toyo Keizai Inc. 
This data source provides lists of shareholders for companies listed on
the stock market or on the over-the-counter market.
The lengths of the shareholder lists vary with the companies.
The data before 2000 contain information on the top 20 shareholders for each company.
On the other hand, the data for 2002 and 2003 contain information on the top 30 shareholders
for each company.
Therefore to uniformly analyze the data we consider the top 20 shareholders for each company.

Types of shareholders include listed companies,
non-listed financial institutions (commercial banks, trust banks,
and insurance companies), officers, and other individuals.
In this article, we don't consider officers and other individuals, so
the shareholding networks are constructed only from companies.
The number of nodes, $N$, and the total number of edges, $K$, vary with the years,
and these are summarized in Table.~\ref{tab:1}.
 
\begin{table}[t]
\centering
\caption{Change in the size of shareholding network $N$, the total number of edges $K$,
and the exponent $\gamma$ of the outgoing degree distribution
$p(k_\textrm{\scriptsize out})\propto k_\textrm{\scriptsize out}^{-\gamma}$}
\label{tab:1}
\begin{tabular}{|c||c|c|c|c|c|c|}\hline
Year&1985&1990&1995&2000&2002&2003\\ \hline\hline
$N$&2,078&2,466&3,006&3,527&3,727&3,770 \\ \hline
$K$&23,916&29,054&33,860&32,586&30,000&26,407 \\ \hline
$\gamma$&1.68&1.67&1.72&1.77&1.82&1.86 \\\hline
\end{tabular}
\end{table}

This paper is organized as follows. In Sec.~\ref{sec:2} we consider the degree distribution
for outgoing edges and show that the outgoing degree
distribution follows a power law function with an exponential cutoff.
In addition, we show that the exponent depends on the year and the country,
while the power law shape is maintained universally.
We also discuss correlations between the exponent
and the long-term shareholding rate and the cross-shareholding rate.
From this examination, we show that the exponent strongly correlates with these quantities.
The last section is devoted to a summary and discussion.

\section{Change of outgoing degree distribution}
\label{sec:2}
If we draw arrows from shareholders to stock corporations, we can represent a
shareholding network as a directed graph.
If we count the number of incoming edges and that of outgoing edges for each node,
we can obtain the degree distribution for incoming degree, $k_\textrm{\scriptsize in}$,
and that for outgoing degree, $k_\textrm{\scriptsize out}$.
However, as explained in Sec.~\ref{sec:1}, the lengths of the shareholder lists vary with the
companies, and thus we consider only the top 20 shareholders for consistency.
Therefore, the incoming degree has an upper bound,
$k_\textrm{\scriptsize in}\leq 20$, while the outgoing degree has no bound.

\begin{figure}[t]
\centering
\begin{minipage}{1.\linewidth}
\includegraphics[width=\linewidth]{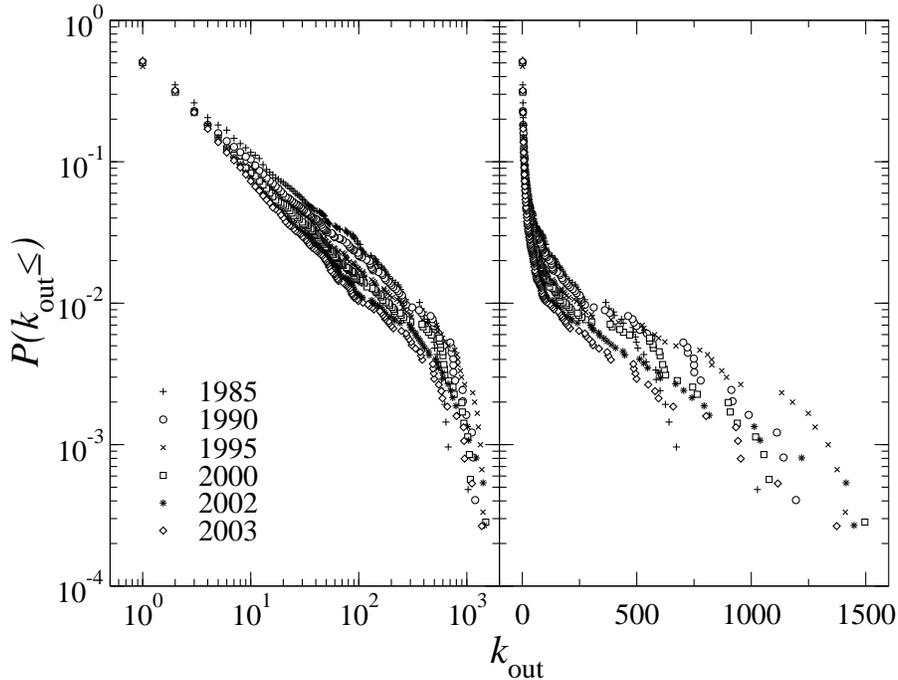}
\end{minipage}
\caption{Log-log plot (left) and semi-log plot (right) of the cumulative probability distribution,
$P(k_\textrm{\scriptsize out}\leq)$, of the outgoing degree $k_\textrm{\scriptsize out}$.
}
\label{fig:degree}       
\end{figure}

The log-log plot of $k_\textrm{\scriptsize out}$ is shown in the left
panel of Fig.~\ref{fig:degree}.
In this figure, the horizontal axis corresponds to $k_\textrm{\scriptsize out}$, and
the vertical axis corresponds to the cumulative probability distribution $P(k_\textrm{\scriptsize out}\leq)$
that is defined by the probability distribution function $p(k_\textrm{\scriptsize out})$,
\[
P(k_\textrm{\scriptsize out}\leq)=\int_{k_\textrm{\tiny out}}^\infty\!\!\!\!\!dk'_\textrm{\scriptsize out}
\;p(k'_\textrm{\scriptsize out}),
\]
in the continuous case.
We can see that the distribution follows the power law function,
$p(k_\textrm{\scriptsize out})\propto k_\textrm{\scriptsize out}^{-\gamma}$, except for the tail part.
The exponent $\gamma$ depends on the year, as summarized in Table.~\ref{tab:1}.
It has also been reported that the degree distributions of shareholding networks for companies listed on
the  Italian stock market (Milano Italia Borsa; MIB), 
the New York Stock Exchange (NYSE), and
the National Association of Security Dealers Automated Quotations (NASDAQ) each
follow the power law distribution \cite{gbcsc2003}.
The exponents are $\gamma_\textrm{\tiny MIB}=1.97$ in 2002,
$\gamma_\textrm{\tiny NYSE}=1.37$ in 2000, and
$\gamma_\textrm{\tiny NASDAQ}=1.22$ in 2000. These are not so different from
the Japanese case.

The semi-log plot is shown in the right panel of Fig.~\ref{fig:degree}, and
the meaning of the axes is the same as in the left panel.
We can see that the tail part of the distribution follows approximately
the exponential function.
The exponential part of the distribution is mainly constructed from financial institutions.
On the other hand, almost all of the power law part of the distribution is constructed
from non-financial institutions.
The above results suggest that different mechanisms work in each
range of the distribution, and some of the reasons for the emergence of this distribution are discussed
in Ref.~\cite{sfa2005b}.

It is reasonable to assume that the change in the exponent $\gamma$ can be attributed to
the change in the pattern of shareholding.
In Japan, since 1987,
a long-term shareholding rate and a cross-shareholding rate have been reported by Nippon Life Insurance (NLI)
Research Institute \cite{NLI2004}.

\begin{figure}[t]
\centering
\begin{minipage}{.85\linewidth}
\includegraphics[width=\linewidth]{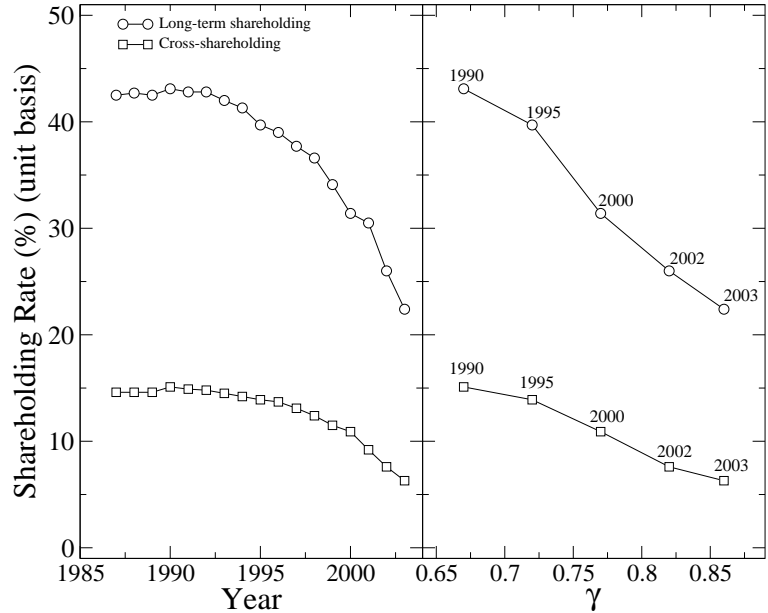}
\end{minipage}
\caption{Change in the long-term shareholding rate and that in the cross-shareholding rate (left),
and the correlations between these rates and the exponent $\gamma$ (right).
}
\label{fig:2}       
\end{figure}

The changes in these rates are shown in the left panel of Fig.~\ref{fig:2}.
In this figure, the horizontal axis corresponds to the year, and
the vertical axis corresponds to the shareholding rate calculated on the basis of
number of shares.
The open circles corresponds to long-term shareholding, and the open squares corresponds
to cross-shareholding.
We can see that both the long-term shareholding rate and the cross-shareholding rate decrease
after 1990.

Correlations of the exponent
with the long-term shareholding rate and with the cross-shareholding rate are shown in the right panel
of Fig.~\ref{fig:2}.
In this figure, the horizontal axis corresponds to the exponent $\gamma$, and
the vertical axis, the open circle, and the open square are the same as in the left panel.
We can see that the exponent has strong and negative correlations with both the long-term shareholding rate and
the cross-shareholding rate.

\section{Summary}
\label{sec:3}
In this article, we considered Japanese shareholding networks as they existed
in 1985, 1990, 1995, 2000, 2002, and 2003.
These networks were represented as a directed graph by drawing arrows from shareholders
to stock corporations.
For these directed shareholding networks, it was shown that the outgoing degree distribution
for each year can be well explained by the power law distribution, except for the tail part.
The exponent depends on the year and the country, while the power law shape is maintained universally.
We also showed that the exponent has strong and negative correlation with both the long-term shareholding
rate and the cross-shareholding rate.
This means that the dissolution of long-term shareholding and cross-shareholding
causes the exponent to increase.

%
%

%

\begin{thebibliography}{99.}
%
%
%





\bibitem{barabasi2002}
Barab\'{a}si AL (2002)
Linked: The New Science of Networks.
Perseus Press, Cambridge, MA

\bibitem{bbw2003}
Battiston S, Bonabeau E, Weisbuch G
(2003) Decision making dynamics in corporate boards.
Physica A 322: 567--582

\bibitem{dm2003}
Dorogovtsev SN, Mendes JFF (2003)
Evolution of Networks: From Biological Nets to the Internet and WWW.
Oxford University Press, Oxford

\bibitem{dyb2003}
Davis G, Yoo M, Baker WE
(2003) The small world of the American corporate elite, 1982--2001.
Strategic Organization 3: 301--326


\bibitem{gbcsc2003}
Garlaschelli D, et~al.
(2003) The scale-free topology of market investments.
to be published in Physica A. arXiv:cond-mat/0310503

\bibitem{ljc2003}
Li X, Jin YY, Chen G
(2003) Complexity and synchronization of the World trade Web.
Physica A 328: 287--296

\bibitem{ljc2004}
Li X, Jin YY, Chen G
(2004) On the topology of the world exchange arrangements web.
Physica A 343: 573--582

\bibitem{NLI2004}
NLI Research Institute Financial Research Group
(2004) The fiscal 2003 survey of cross-shareholding.
http://www.nli-research.co.jp/index.html 

\bibitem{sfa2003}
Souma W, Fujiwara Y, Aoyama H
(2003) Complex networks and economics.
Physica A 324: 396--401

\bibitem{sfa2004}
Souma W, Fujiwara Y, Aoyama H
(2004) Random matrix approach to shareholding networks.
Physica A 344: 73--76

\bibitem{sfa2005a}
Souma W, Fujiwara Y, Aoyama H
(2005) Heterogeneous economic networks. In: Namatame A, et~al. (eds)
the proceedings of the 9th Workshop on Economics
and Heterogeneous Interacting Agents.
Springer-Verlag, Tokyo, to be published. arXiv:physics/0502005

\bibitem{sfa2005b}
Souma W, Fujiwara Y, Aoyama H
(2005) Shareholding networks in Japan In: Mendes JFF, et~al. (eds)
the proceedings of the International Conference ``Science of Complex Networks:
from Biology to the Internet and WWW".
Springer-Verlag, Berlin, to be published. arXiv:physics/0503177





























\end{thebibliography}
%



\printindex
\end{document}